\documentclass[]{aa}
\usepackage{psfig,epsfig,graphicx}
\usepackage{amsmath}
\usepackage{natbib}
\bibpunct{(}{)}{;}{a}{}{,}

\begin{document}


\title{On the contribution of microlensing to X-ray variability of
high-redshifted QSOs}
\author{A.F. Zakharov\inst{1,2,3,4}, L. \v C. Popovi\'c\inst{5,6,7},
P. Jovanovi\'c\inst{5,7}}

\offprints{Popovi\'c L., \email{lpopovic@aip.de}} \institute{
National Astronomical Observatories, Chinese Academy of Sciences,
100012, Beijing, China, \and Institute of Theoretical and
Experimental Physics,
           25, B.Cheremushkinskaya st., Moscow, 117259, Russia
 \and
 Astro Space
Centre of Lebedev Physics Institute, Moscow
\and Isaac Newton
Institute of Chile,
 Moscow Branch
 \and Astronomical Observatory, Volgina 7, 11160 Beograd, Serbia
\and Astrophysikalisches Institut Potsdam, An der Sternwarte 16,
14482 Potsdam, Germany \and Isaac Newton Institute of Chile,
 Yugoslavia Branch
 }
\authorrunning{Zakharov, Popovi\'c \& Jovanovi\'c}
\titlerunning{On the contribution of microlensing to X-ray variability
of 
QSOs}
\date{Received / accepted }

\abstract{We consider a contribution of microlensing to the X-ray
variability of high-redshifted QSOs. Such an effect could be
caused by stellar mass objects (SMO) located in a bulge or/and in
a halo of this quasar as well as  at cosmological distances
between an observer and a quasar. Here, we  not consider
microlensing caused by deflectors in our Galaxy  since it is
well-known from recent MACHO, EROS and  OGLE observations that the
corresponding optical depth for the Galactic halo and the Galactic
bulge is lower than $10^{-6}$. Cosmologically distributed
gravitational microlenses could be localized in galaxies (or even
in bulge or halo of  gravitational macrolenses) or could be
distributed in a uniform way. We have analyzed both cases of such
distributions. As a result of our analysis, we obtained that the
optical depth for microlensing caused by stellar mass objects is
usually small for quasar bulge and  quasar halo gravitational
microlens distributions ($\tau\sim 10^{-4}$). On the other hand,
the optical depth for gravitational microlensing caused by
cosmologically distributed deflectors could be significant and
could reach $10^{-2} - 0.1$ at $z\sim 2$.  This means that
cosmologically distributed deflectors may contribute
significantlly  to the X-ray variability of high-redshifted QSOs
($z>2$). Considering that the upper limit of the optical depth
($\tau\sim 0.1$) corresponds to the case where dark matter forms
cosmologically distributed deflectors,   observations of the X-ray
variations of unlensed QSOs can be used for the estimation of the
dark matter fraction of microlenses.

 }

\maketitle

\section{Introduction}

The X-ray radiation of Active Galactic Nuclei (AGNs), in the
continuum as well as in
spectral lines, has rapid and irregular variability
(see e.g. \cite{Mar81,Bar86,Law93,Gre93,Tur99,Wea01,Man02}, etc.).
 X-ray flux variability has long been known to be a common property of
active galactic nuclei (AGNs), e.g. Ariel 5 and HEAO 1 first
revealed long-term (days to years) variability in AGNs  and
by uninterrupted observations of EXOSAT  rapid (thousands of
seconds) variability was also established as common in these
sources (see, for example reviews by \cite{Mush93,Ulrich97} and
references therein). X-ray flux variations are observed on
timescales from $\sim$1000 s to years, and amplitude variations of
up to an order of magnitude are observed in the $\sim$ 0.1 - 10
keV band. It was first suggested by Barr \& Mushotzky (1986)
 that the flux variation of an AGN is inversely
proportional to its luminosity.
\cite{Law93} and  \cite{Gre93} confirmed the
variability-luminosity relationship, finding that the variability
amplitude ($\sigma$) varies with luminosity as
$\sigma=L_X^{-\beta}$ with $\beta\approx0.3$. Recently, \cite{Man02}
 analyzed the variability of a sample of 156 radio-quiet quasars
taken from the ROSAT archive, considering the trends in variability of the
amplitude with luminosity and with redshift. They found that there
was evidences for a growth in AGN X-ray variability amplitude
towards high redshift ($z$) in the sense that AGNs of the same
X-ray luminosity were more variable at $z>2$. They explained the
$\sigma$ {\it vs.} $z$ trend assuming that the high-redshifted
AGNs accreted at a larger fraction of the Eddington limit than
the low-redshifted ones.

On the other hand, the contribution of microlensing to AGN
variability was considered in many papers (see e.g.
\cite{Hawk93,Hawk02, Wamb01,Wamb01b,Zakh97}, and references
therein). Moreover, recently X-ray microlensing of AGN has been
considered \citep{Popov01a,Tak01,Chart02a,Popovic03,Dai03}. Taking into
account that the X-rays of AGNs are generated in the innermost and
very compact region of an accretion disc, the X-ray radiation in
the continuum as well as in a line can be strongly affected by
microlensing \citep{Popovic03}.\footnote{Simulations of X-ray
line profiles are presented in a number of papers, see, for
example, \cite{Zak_rep02,Zak_rep02a,Zak_rep02_xeus,Zak_rep03_ASR,Zak_rep03_Su}
and references therein, in particular \cite{ZKLR02} showed that an
information about magnetic filed may be extracted from X-ray line
shape analysis; \cite{Zak_rep03_AA} discussed signatures of X-ray
line shapes for highly inclined accretion disks.}

 Recent
observations of three lens systems seem to support this idea
\citep{Osh01,Chart02a,Dai03}.  \cite{Popovic03}  showed that
objects in a foreground galaxy with very small masses can cause
strong changes in the X-ray line profile.  This  fact may
indicate  that the observational probability of X-ray variation
due to microlensing events is higher  than in
the UV and optical radiation of AGNs. It is connected with
the fact that typical sizes of X-ray emission regions are much
smaller than typical sizes of those producing optical and UV bands.
Typical optical and UV emission region sizes could be
comparable or even larger than Einstein radii of microlenses and
therefore microlenses magnify  only a small part of the region emitting in
the optical or UV band (see e.g. \cite{Pop01b,Aba02}, for UV and
optical spectral
line
region). This is reason that it could be a
very tiny effect from an observer point of view.

The aim of this paper is to discuss the contribution of
microlensing to the relation $\sigma$ {\it vs.} $z$ for X-ray radiation
considering
the recent results given by \cite{Man02} and \cite{Popovic03}. In
the next section we will consider the optical depth.

\section{The optical depth}

The optical depth $\tau$ (the chance of seeing a microlens (ML))
is the probability that at any instant of time a source is covered
by the Einstein ring of a deflector. Here we will consider
deflectors from the host bulge and halo as well as at
cosmological distances between an observer and  source. We will
not consider microlensing   caused by Galactic microlenses
since it is well-known from recent MACHO, EROS and OGLE
observations that the corresponding optical depth for Galactic halo
and Galactic bulge is lower than $10^{-6}$. { Therefore, by analogy, one
could expect that the optical depth for microlensing due to
objects in the halo or/and bulge of a quasar is small (similar to
the optical depth for microlensing in Galaxy). However, it would be
appropriate to present some more accurate estimates for optical depths for
microlensing by bulge/halo objects
assuming reasonable density distribution
models of QSOs. The reason for this is that, as we mentioned above,
the X-ray emission regions are much smaller than UV/optical ones, and
 cosmologically distributed  small mass deflectors or deflectors
from QSO bulge/halo can produce significant magnification in X-ray radiation,
but it will
not happen in the UV/optical band.
 Below we will demonstrate such
variations
of parameters that could cause some rise of optical depth in bulge/halo of
a
QSO.}
\subsection{Quasar Bulge microlenses}

In this section we consider gravitational microlensing  caused by stellar
mass objects in the bulge of an observed quasar. Of
course, to calculate an optical depth we have to know the radial
mass density distribution in the QSO bulge. In this case the optical
depth could be evaluated by the integral
\begin {eqnarray}
 \tau \sim \frac{4\pi G}{c^2} \int_0^R \rho(r) r
d\,r, \label{eq_bulge1}
\end {eqnarray}
where $R$ is the bulge radius. For qualitative discussions of the
optical depth range we make the assumption of
 constant mass density (see also
\cite{Popovic03}).  Evaluating this integral, we obtain
\begin {eqnarray}
 \tau \sim \frac{2\pi G}{c^2} \rho_0 R^2=
 \frac{3 G}{2c^2} \frac{M_{\rm bulge}}{R},
  \label{eq_bulge2}
\end {eqnarray}
where $\rho_0=\dfrac{3 M_{\rm bulge}}{4 \pi R^3}$ is the average
density of the bulge. It is clear that the maximal optical depth
corresponds to the most compact galactic bulge for a fixed bulge
mass.
For an estimate of the bulge mass use can be made of scaling from the
black hole mass; \citep{Mclure02} give $M_{\rm bh}=0.0012M_{\rm bulge}$,
and
\citep{Shields02} $M_{\rm bulge}=10^{2.8} M_{\rm
bh}$. However, for Seyfert 1 galaxies ratios of
the central black hole mass and the bulge mass could be about
$1 \times 10^{-4}$ \citep{Bian03}.
We can derive an upper limit from the estimate by
 \cite{Czerny01} that for
 AGN  $M_{\rm bulge}\approx 10^{12}
M_\odot$ (with  $M_{AGN}\sim 10^{13}M_\odot$ \citep{Shields02}).
\cite{Schade00} found typical values for the radii of AGN bulges
in the range 1--10 kpc. So, using the lower limit for the AGN
bulge radius and the total mass estimation $M_{tot}$, we obtain an
upper limit of the optical depth for microlensing by bulge stellar
mass objects of $\tau_{\rm bulge} \sim 3.5 \times 10^{-5}$. This
upper limit is about the value evaluated earlier \citep{Popovic03}
and the contribution to the total optical depth for microlensing
is small. Microlensing would thus be detectable only in a small
fraction of quasars.

\subsection{Quasar Halo microlenses}

\subsubsection{Singular isothermal sphere model}

Here we assume  a mass density distribution  described
by a singular isothermal sphere model, namely
\begin {eqnarray}
\rho (y)=\left\{
\begin{array}{ll}
\dfrac{\rho_0 r}{y^2}, & r \leq  y  \leq R,\\
0, & y  > R {\rm ~or~} y < r,
\end{array}
\right.,
\end {eqnarray}
where $r$ is the inner and $R$ is the outer radius of halo, and
 $\rho_0$ is the mass density at the inner radius $r$,
\begin {eqnarray}
 \tau_{\rm halo} \sim \frac{4\pi G}{c^2} \int_r^R \rho \frac{D_d
 (D_s-D_d)}{D_s} d D_d, \label{eq_halo1}
\end {eqnarray}
Evaluating this integral, we obtain
\begin {eqnarray}
 \tau_{\rm halo} \sim \frac{4\pi G}{c^2}\rho_0 r^2 \ln \frac{R}{r}.
 \label{eq_halo2}
\end {eqnarray}

The halo mass can be expressed as
\begin {eqnarray}
M_{\rm halo} = \int_r^R  \frac{\rho_0 r^2}{y^2} 4 \pi y^2 dy =
4\pi \rho_0 r^2R.
 \label{eq_halo3}
\end {eqnarray}
Thus,
\begin {eqnarray}
\rho_0 =   \frac{M_{\rm halo}}{4 \pi r^2 R}
 \label{eq_halo4}
\end {eqnarray}
and
\begin {eqnarray}
 \tau_{\rm halo} \sim \frac{G}{c^2}\frac{M_{\rm halo}}{R} \ln \frac{R}{r}.
 \label{eq_halo5}
\end {eqnarray}
Typical halo masses are in the range $10^{11}-10^{14}M_\odot$
range \citep{Bullock01,Ferrarese02} and typical halo radii are $R$ are
$\sim$  few $\times 10^{2}$ kpc \citep{Klypin02,Ferrarese02},
and typical inner radii $r \sim$  a few $\times 10$ kpc
\citep{Ferrarese02}, we can estimate the optical depth using
these values. Assuming that $M_{\rm halo} =
10^{14}M_\odot$,  $R \sim 10^{2}$ kpc and $r \sim 5$ kpc we   obtain
$\tau_{\rm halo} \sim \tau_{\rm bulge} \sim
7\cdot 10^{-5}$.

\subsubsection{Navarro -- Frenk -- White halo (NFW) profiles}

Let us calculate the optical depth for Navarro -- Frenk -- White
(NFW)  halo profiles of mass density distributions. A
two-parameter form for halo profiles was proposed by
\cite{NFW95,NFW96,NFW97}
\begin {eqnarray}
\rho_{\rm NFW} (r)=\dfrac{\rho_s}{(r/r_s)(1+r/r_s)^2},
\label{eq_halo6}
\end {eqnarray}
where $r_s$ is a characteristic inner radius and $\rho_s$ is the
corresponding inner density, $\rho_s=4 \rho_{\rm NFW}(r_s)$ and
$\rho_s=\rho_{\rm NFW}(0.466 r_s)$ \citep{Bullock01}, where $0.466
r_s$ is the approximate solution of the equation
\begin {eqnarray}
(r/r_s)^3+2(r/r_s)^2+(r/r_s)-1=0. \label{eq_halo6a}
\end {eqnarray}

 \cite{NFW95,NFW96,NFW97} showed that these halo profiles provide
a good fit over a large range of masses and for several cosmological
scenarios (including a flat cosmological model with $\Omega_m=0.3$
and $\Omega_\Lambda=0.7$). \cite{Bullock01} confirmed the success of
this model at $z=0$, but mentioned that the NFW model significantly
over-predicts the concentration of halos at early times $z >
 1$ and suggested some modifications of the NFW model. However,
we will use the standard NFW model.

One can calculate the halo mass
\begin {eqnarray}
M_{\rm halo} = 4\pi \rho_s r_s^3 A(c_{\rm vir}),
 \label{eq_halo7}
\end {eqnarray}
where
\begin {eqnarray}
 A(c_{\rm vir})=\ln(1+c_{\rm vir})- \frac{1+c_{\rm vir}}{c_{\rm vir}}
 \label{eq_halo8}
\end {eqnarray}
and $c_{\rm vir}=R/r_s$. Using Eq. (\ref{eq_halo1}) and the NFW halo
profile, one obtains
\begin {eqnarray}
 \tau_{\rm halo} \sim \frac{2 \pi G}{c^2} \rho_s r_s^2
 \label{eq_halo9}
\end {eqnarray}
and substituting  $\rho_s$ from Eq.(\ref{eq_halo7})
\begin {eqnarray}
\rho_{\rm halo} =   \frac{M_{\rm halo}}{4 \pi r_s^2 A(c_{\rm
vir})}
 \label{eq_halo10}
\end {eqnarray}
we obtain
\begin {eqnarray}
 \tau_{\rm halo} \sim \frac{G}{c^2}\frac{M_{\rm halo}}{2 r_s A(c_{\rm vir})}.
 \label{eq_halo11}
\end {eqnarray}
Since typical $c_{\rm vir}$ values are in the range 5--30,
$A(c_{\rm vir})$  varies in the range 1--3, and $r_s \sim $  a
few $\times 10$ kpc \citep{Ferrarese02}. Assuming $M=10^{14}
\times M_\odot$, $r_s= 3$~kpc, $A(c_{\rm vir})=2$, we obtained
$\tau_{\rm halo} \sim 4 \times 10^{-4}$. Therefore, the optical depth
estimates by \cite{Popovic03} are realistic if we consider
 objects inside the halo and/or
bulge. We recall that they found the optical depth to be in the range
 $10^{-4} - 10^{-3}$.

\subsection{Cosmological distribution of microlenses}

To estimate the optical depth we will use the point size source
approximation for an emitting region of X-ray radiation. It means
that the size of emitting region  is smaller than this
Einstein --
Chwolson radius. This approximation is used commonly to
investigate microlensing in optical and UV bands. The typical
 Einstein -- Chwolson radius of a lens can be expressed in
the following way \citep{Wamb01}
\begin {eqnarray}
r_{\rm EC}=\sqrt{\frac{4GM}{c^2}\frac{D_s D_{ls}}{D_l}} \sim 4
\times 10^{16} \sqrt{M/M_\odot}~{\rm cm},
 \label{eq_cosmol_l1}
\end {eqnarray}
where "typical" lens and source redshift of $z \sim 0.5$ and $z
\sim 2$ were chosen, $M$ is the lens mass, $D_l$, $D_s$ and
$D_{ls}$ are angular diameter distances between observer and lens,
observer and source, lens and source respectively. A typical
quasar size is parameterized in units of $10^{15}$ cm \citep{Wamb01}.
Since the point size source approximation for an emitting region
is reasonable for optical and for UV bands, and as it is generally adopted
that X-ray radiation is formed in the inner parts of accretion
disks we can use this an approximation for X-ray sources.
However, let us make some estimates.  The relevant length scale
for microlensing in the source plane for this sample
\begin {eqnarray}
R_{\rm EC}=r_{\rm EC}\frac{D_s}{D_l} \sim 1 \times 10^{17}~{\rm
cm}.
 \label{eq_cosmol_l2}
\end {eqnarray}

Even if we consider a supermassive black hole in the center of the
quasar $M_{\rm SMBH}= 10^{9} M_\odot$, then its Schwarzschild
radius is $r_g=3\times 10^{14}$ cm and assuming that the emission
region for the X-ray radiation is located near the black hole
$r_{\rm emission} < 100\,r_g =3\times 10^{16}$~cm, we obtain that
$r_{\rm emission} < R_{\rm EC}$, therefore the point size source
approximation can be adopted for the X-ray emitting
region.\footnote{For example, \cite{Chart02a} found evidence for
X-ray microlensing in the gravitationally lensed quasar MG
J0414+0534 ($z=2.639$), where according to their estimates $M_{\rm
SMBH}$ is in the range $3.6 \times 10^{6}(\beta/0.2)^2$ and $1.1
\times 10^{7}(\beta/0.2)^2 M_\odot$ ($\beta \sim 1$).
Therefore a typical emission region is much smaller than the
Einstein -- Chwolson radius $R_{\rm EC}$, since following
\cite{Chart02a} one could assume that the emitting region corresponds
to $(10-1000)\,r_g$ or $\sim 1.5 \times 10^{14} - 1.5 \times
10^{16}$~cm for a $10^{8}M_\odot$ black hole.} Note that sometimes this
approximation cannot be used
when the microlens lies in the bulge or halo of a quasar (see
previous subsections), because then  the Einstein --
Chwolson radius would be about several astronomical units, since
we have $D_l \sim D_s$, $D_{ls} << D_s$ from Eqs.
(\ref{eq_cosmol_l1},\ref{eq_cosmol_l2}),
\begin {eqnarray}
R_{\rm EC} \sim r_{\rm EC}, \nonumber
\end {eqnarray}
and
\begin {eqnarray}
R_{\rm EC} \sim 9\,{\rm au}
\left(\frac{M}{M_\odot}\right)^{1/2}\left(\frac{D_{ls}}{10\,{\rm
kpc}}\right)^{1/2}  \sim 10^{14}\,{\rm cm}.
 \label{eq_cosmol_l3}
\end {eqnarray}
In this case one has to  take into account the size of the X-ray emission
region.

To  evaluate the optical depth, we assume  a source
located at
redshift
$z$.

The expression for optical depth  has been taken  from
\cite{Wan96,TOG84,Fuk91}
\begin {eqnarray}
\tau^p_L= \frac{3}{2}\frac{\Omega_L}{\lambda(z)}
\int_0^z dw
\frac{(1+w)^3[\lambda(z)-\lambda(w)]\lambda(w)}
       {\sqrt{\Omega_0(1+w)^3+\Omega_\Lambda}},
       \label{eq_cosmol2}
\end {eqnarray}
where $\Omega_L$ is the matter fraction in compact lenses,
\begin {eqnarray}
\lambda(z)= \int_0^z
\frac{dw}{(1+w)^2\sqrt{\Omega_0(1+w)^3+\Omega_\Lambda}},
\label{eq_cosmol3}
\end {eqnarray}
is the affine distance (in units of $cH^{-1}_0)$.

\begin{figure}[h!]
\includegraphics[width=7.5cm]{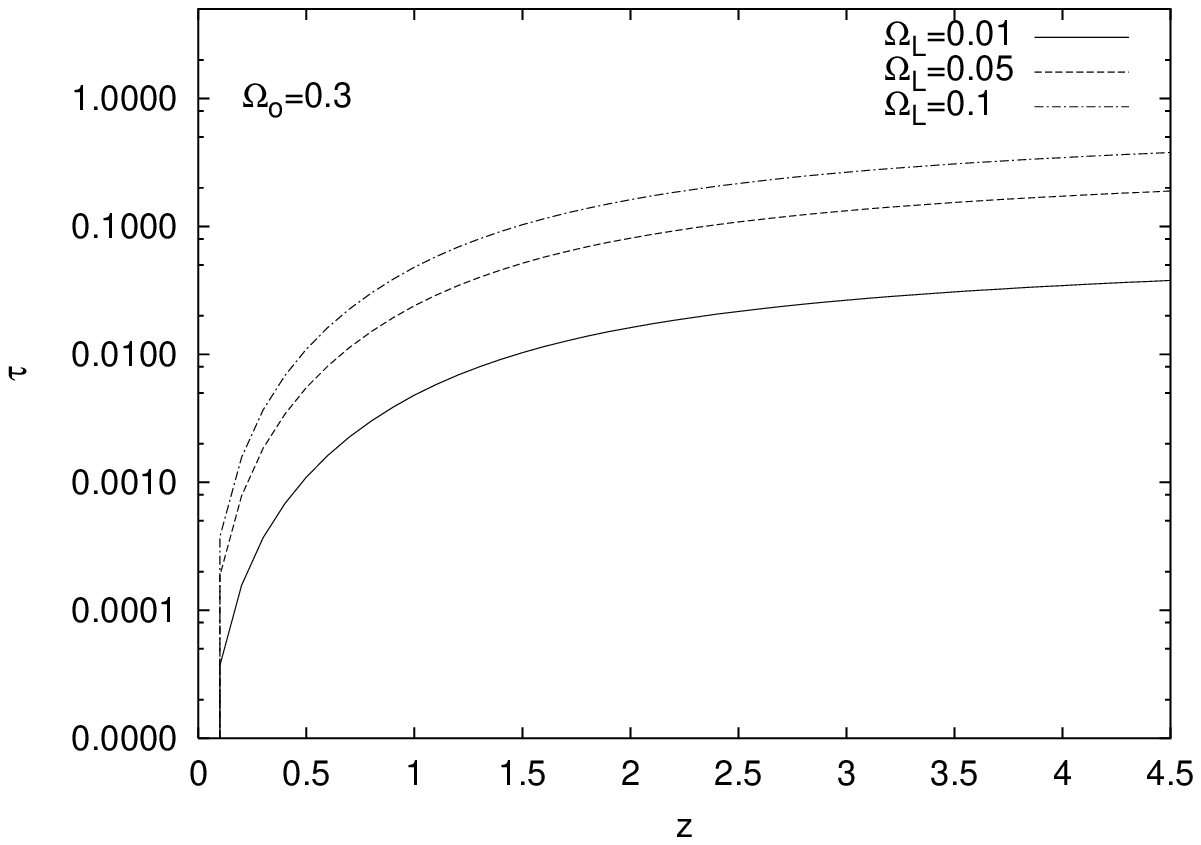}
\includegraphics[width=7.5cm]{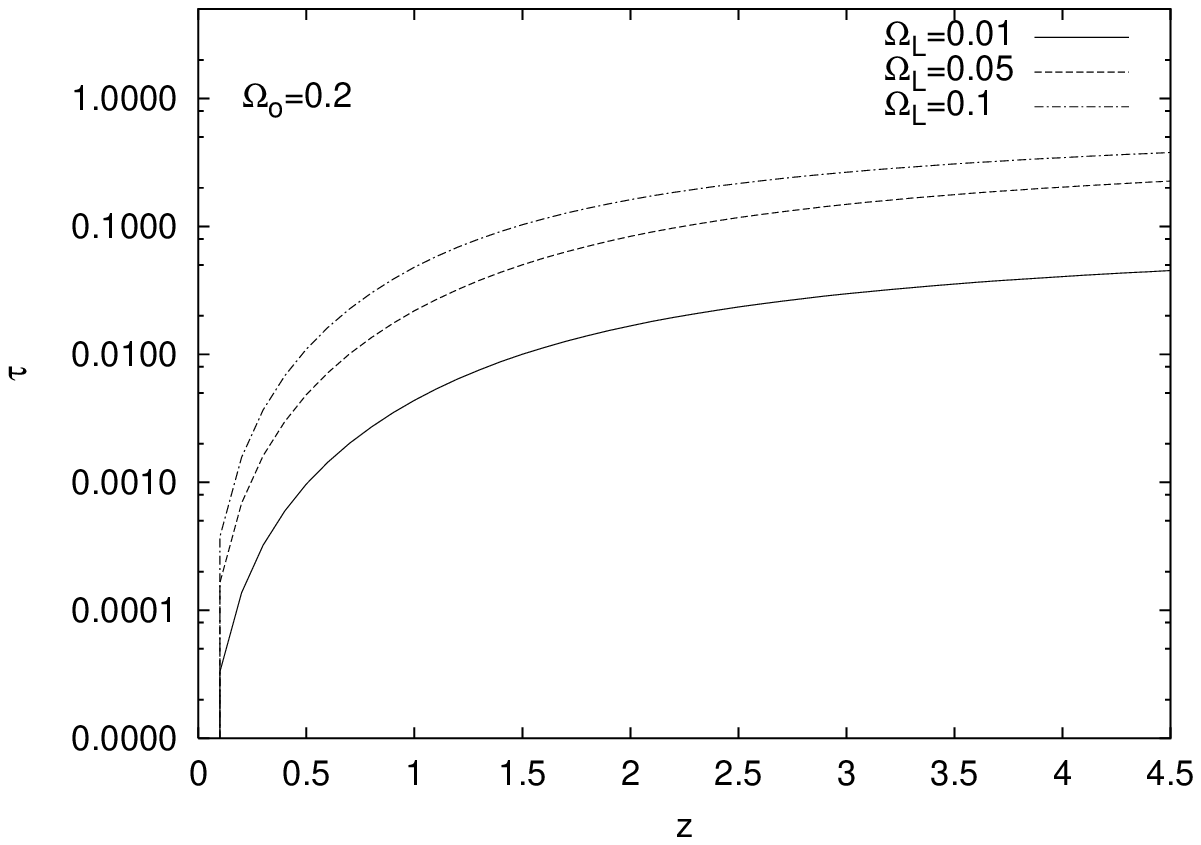}
\includegraphics[width=7.5cm]{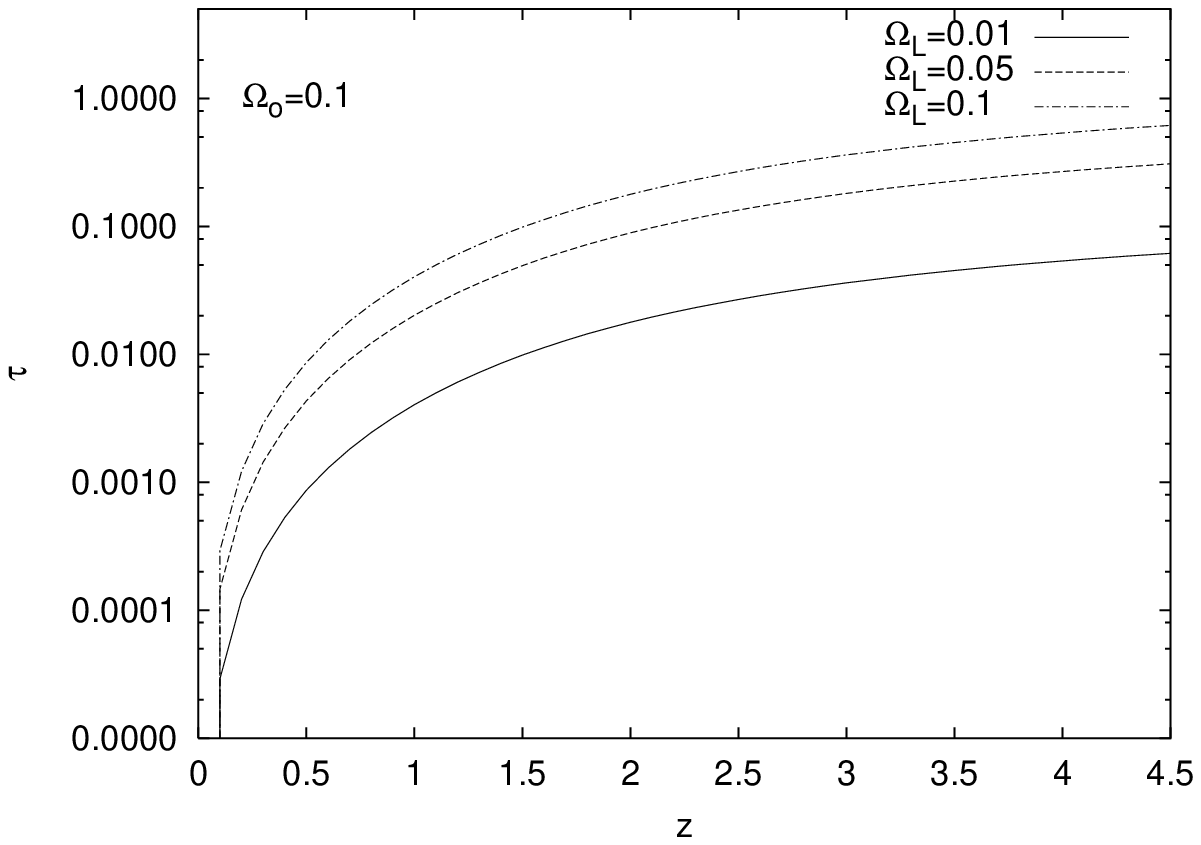}
\caption{The calculated optical depth as a function of redshift
for different value $\Omega_L$ and $\Omega_0$.}
 \label{fig1}
\end{figure}

We will use some realistic cosmological parameters to evaluate the
integral (\ref{eq_cosmol2}). According to the cosmological SN
(Supernova) Ia data
 and  cosmic microwave background (CMB) anisotropy one can take
$\Omega_\Lambda \approx 0.7,
\Omega_0  \approx 0.3$
\citep{Per99,Bon01,Bal01,Lahav02,Peebles02}. Recent CMB anisotropy
observations by the WMAP satellite team have confirmed important
aspects of the current standard cosmological model,  the
WMAP team determined  $\Omega_\Lambda \approx 0.73, \Omega_0
\approx 0.27$ \citep{Ben03,Sper03} for the "best" fit of
cosmological parameters (see also  \cite{Bridle03} for
discussion). Therefore we will assume  $\Omega_0 = 0.3$ and
$\Omega_0 = 0.2$ as realistic cases. If we assume that
microlensing is caused by stars we have to  take into account
cosmological constraints  on baryon density.  Big Bang
Nucleosynthesis (BBN) calculations together with observational
data about the abundance of $^2$D give the following constraints on
the cosmic baryon density \citep{Omeara01,Burles01,Turner02}
\begin {eqnarray}
\Omega_b h^2 = 0.02 \pm 0.002,
\label{eq_cosmol4}
\end {eqnarray}
 taking into account the Hubble constant estimation $h=0.72\pm
0.08$ \citep{Freedman01}. However,
\cite{Parodi00,Tammann02,Tammann02a} give lower limits for
$h=0.585 \pm 0.063$. Therefore, using for example the estimate by
\citet{Freedman01} one could obtain for the cosmic baryon density
\citep{Turner02}
\begin {eqnarray}
\Omega_b  = 0.039 \pm 0.0075. \label{eq_cosmol5}
\end {eqnarray}
Using CMB anisotropy data of
the BOOMERANG and MAXIMA-1 experiments \cite{Stompor01} found that
\begin {eqnarray}
\Omega_b h^2 = 0.033 \pm 0.013 \label{eq_cosmol6}.
\end {eqnarray}
An analysis of recent WMAP data on CMB anisotropy gives as the
best fit \citep{Sper03}
\begin {eqnarray}
\Omega_b h^2 = 0.0224 \pm 0.0009, \label{eq_cosmol7}
\end {eqnarray}
which is very close  to the BBN constraints, but with much
smaller error bars.

\begin{table}
\begin{center}
\caption[]{The calculated optical depth as a function of redshift
for different values of $\Omega_L$ and $\Omega_0=0.3$.}
\begin{tabular}{|c|c|c|c|}
\hline
$z\backslash\Omega_L$ & 0.01 & 0.05 & 0.10 \\
\hline \hline
0.5 & 0.001100 & 0.005499 & 0.010998 \\
1.0 & 0.004793 & 0.023967 & 0.047934 \\
1.5 & 0.010310 & 0.051550 & 0.103100 \\
2.0 & 0.016196 & 0.080980 & 0.161959 \\
2.5 & 0.021667 & 0.108334 & 0.216669 \\
3.0 & 0.026518 & 0.132590 & 0.265180 \\
3.5 & 0.030770 & 0.153852 & 0.307703 \\
4.0 & 0.034504 & 0.172521 & 0.345042 \\
4.5 & 0.037804 & 0.189018 & 0.378037 \\
5.0 & 0.040742 & 0.203712 & 0.407424 \\
\hline
\end{tabular}
\end{center}
\label{tabl1}
\end{table}

 Therefore, the cases with $\Omega_0=0.3$ and $\Omega_L =
0.05$ ($\Omega_L = 0.01$) can be adopted as realistic (the top
panel in Fig. \ref{fig1}, here we assume that almost all baryon
matter { and a small fraction of non-baryon matter}
can form microlenses ($\Omega_L = 0.05$), or, alternatively, that about
25\% of
baryon matter forms such microlenses ($\Omega_L = 0.01$)).
However, for both  cases and for distant objects ($z \sim
2.0$) the optical depth could reach $ \sim 0.01 - 0.1$ (see Table 1
and Fig. 1). If about
30\% of non-baryonic dark matter forms objects with stellar
masses, $\Omega_L = 0.1$ can be adopted, { and
then $\tau\sim 0.2$ at $z\sim 2$}.
The optical depths for  realistic values of $\Omega_L$  as
a function of  redshifts are presented in  Table 1.
 The middle and bottom
panels of Fig. \ref{fig1} show the  optical depth as a function of
redshift for chosen cosmological parameters (densities).

Recently, \cite{Wyithe02} considered probability distributions for
the cases when lensing objects are concentrated in galaxies. The
authors found that about 1\% of high-redshift sources ($z \sim
3$) are microlensed by stars at any  time. The microlensing rate
by stars in elliptical/S0 galaxies is 10 times higher than in
spiral galaxies. Multiple imaged sources dominate the stellar
microlensing statistics. However, if CDM halos are composed of
compact objects, \cite{Wyithe02} concluded that the microlensing rate
should be about 0.1, i.e. $\sim 1$ high-redshift source out of 10 is
microlensed at any  time.

\cite{Wyithe02b} calculated variability rates for a hypothetical
survey. Let us recall  their results. For a limiting quasar
magnitude $m_B=21$  the authors found that the probability that a
quasar could show a variability larger than $m_B=0.5$ due to microlensing
by stars is about
$2 \times 10^{-3}$  (the
cosmological density of stars is assumed to be equal to
$\Omega_*=0.005$). 90\% of these events are in multiple-imaged
systems. Therefore, microlenses in gravitational lenses forming
multiple-imaged quasars dominate in these  statistics.
Assuming that a dark halo (truncated so that the total mass density
equals the critical density) is also composed of compact
objects, the fraction of quasar images which exhibit
microlensing variability  larger than $m_B=0.5$ rises  to $\sim
10\%$.
Thus, \cite{Wyithe02b} pointed out that the comparison of lensed
and un-lensed quasars will provide a powerful test for dark compact
objects in the halo.

\subsection{Microlensing of gravitationally lensed objects}

Just after the discovery of the first multiple-imaged quasar
QSO~0957+561~A,B by \cite{Walsh79} the idea of microlensing by low
mass stars in a heavy halo was suggested by \cite{Gott81}. First
evidence of quasar microlensing was found by \cite{Irwin89}. Now
there is a number of known gravitational lens systems
\citep{Claeskens02,Browne03} and some of them show evidence
for microlensing \citep{Wamb01}.

In this subsection we consider the optical depth for gravitational
microlensing in multiple-imaged quasars. There  is cumbersome approaches
 to calculate probability for this case. See for
example, the papers by
\cite{Deguchi87,Seitz94,Seitz94a,Neindorf03}. Here we will present
some rough estimates for such a phenomenon,  using calculations by
\cite{Turner90,Wan96} for a flat universe with $\Lambda$-term.
According to \cite{Turner90} optical depth for macrolensing
\begin {eqnarray}
\tau_{GL}= \frac{F}{30} \left[\int_1^y \frac{dw}{(\Omega_0
w^3-\Omega_0+1)^{1/2}}\right]^3, \label{eg_gl1}
\end {eqnarray}
where $z_Q=y-1$ ($z_Q$ is the quasar redshift) and
\begin {eqnarray}
F= 16 \pi^3 n_0
\left(\frac{c}{H_0}\right)^3\left(\frac{\sigma}{c}\right)^4,
\label{eg_gl2}
\end {eqnarray}
$F$ characterizes the gravitational lens effectiveness, $\sigma$
is the one-dimensional velocity dispersion and $n_0$ is the co-moving
space density. According to \cite{Turner90,TOG84} the
effectiveness $F$ can be chosen to be $F=0.15$. As  was shown by
\cite{Turner90}, for the most popular cosmological model
$\Omega_0=0.3$ and a distant quasar $z_Q=2$ the optical depth
could be about $0.01$. In  Fig. 2  the optical depth
as a function of  cosmological redshift
is given. As one can see from Fig. 2, the optical depth has similar trend
as in the case of cosmologically distributed objects.

\begin{figure}[h!]
\includegraphics[width=7.5cm]{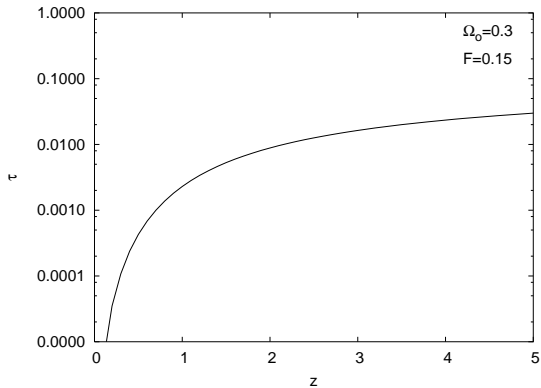}
\caption{The calculated optical depth for gravitational
macrolensing $\tau_{GL}$ as a function of redshift for the most
realistic cosmological matter density $\Omega_0=0.3$.}
\label{fig2}
\end{figure}

If we  try to find the microlensing phenomenon in  multiply imaged
quasars,  we should recall that \cite{Wyithe02b} showed that
if we restrict ourselves to quasars for which the sum of the
macro-images is brighter than $m_B=21$,  one image in three
multiply imaged quasars should vary by more than 0.5 mag during 10
years of monitoring. This means that roughly speaking the probability of
microlensing for multiple imaged quasars is about 0.3.

\subsection{Observed features of microlensing for quasars}

More than 10 years ago \cite{Hawk93} (see also \cite{Hawk96,Hawk02})
 put
forward the idea that nearly all quasars are being microlensed.
 Recently, \cite{Hawk02} considered three basic models to
explain AGN variability: the disc instability model proposed by
\cite{Rees84},
the starburst model developed by \cite{Aret94}
as an alternative, and finally  the idea
that the observed variations are not intrinsic to the AGN, but a
result of gravitational microlensing by stellar mass objects along
the line of sight \citep{Hawk93}. Suggesting that different mechanisms
dominate in different luminosity regimes \cite{Hawk02} divided AGN
into two categories, quasars with $M_B < -23$ and Seyfert galaxies
with $M_B > -23$.

To distinguish different models of variability \cite{Hawk02} used
quantitative predictions for the statistics of AGN variability
based on structure functions of \cite{Kawa98}. \cite{Hawk02} analyzed
about 1500 quasars in the central 19 deg$^2$ of ESO/SERC Field 287 up
to  magnitude  $B_J=22$, and  610 have been confirmed
with
redshifts. The structure function was calculated for a sample of 401
quasars from the survey of \cite{Hawk96}. For
comparison he considered the results of monitoring Seyfert galaxy
NGC~5548 and a sample of 45 Seyfert galaxie from the survey of
\cite{Hawk96}. He calculated structure functions slopes of two
class of AGN and found that the slope is  $0.36\pm 0.02$ for
Seyfert galaxies and $0.2\pm 0.01$ for quasars. Since the model
prescriptions give structure function slopes of $0.83\pm 0.08$
for the starburst model, $0.44\pm 0.03$ for the disc instability
model and $0.25\pm 0.03$ for microlensing,
 the observational results  favor  the disc
instability model for Seyfert galaxies, and the microlensing model
for quasars. The starburst and disc
instability models are ruled out for quasars, while the microlensing model
is
in good agreement  with the observations. As was shown by
\cite{Hawk96} the cosmological density of microlenses should be
comparable with the critical density or at least with $\Omega_m \sim
0.3$. However,  the analysis of the
structural function only cannot confirm or rule out the hypothesis
of microlensing origin of quasar variability, but it is  an additional
argument in favor of the microlensing model.

\section{Discussion}

As  was mentioned earlier by \cite{Popovic03} the probability of
microlensing by stars or other compact objects in halos and bulges of
quasars is very low (about $10^{-4} - 10^{-3}$). However, for
cosmologically distributed microlenses it could reach $10^{-2} - 0.1$ at
$z\sim 2$.
The
upper limit $\tau \sim 0.1$ corresponds to the case where
compact dark matter forms cosmologically distributed microlenses.
As one can see from Fig. 1, in this case the optical
depth for the considered value of $\Omega_0$  is around 0.1  for
 $z > 2$. This indicates that   such a phenomenon could be  observed
frequently, but only for
distant sources ($z \sim 2$). Moreover, it is in good agreement with
the trend in the variability amplitude with redshift found by
\cite{Man02}, where
AGNs of the same X-ray luminosity are more variable at $z>2$.

To investigate  distortions of spectral line shapes due to
microlensing \citep{Popovic03} the most real candidates are
multiply imaged quasars, since the corresponding probability could
be about 0.3 (for magnification of one image $\Delta m =0.5$
during 10 years). However,
these cases the simple point-like microlens model may  not be very good
approximation \citep{Wamb01,Wamb01b} and one should use a numerical
approach, such as the MICROLENS ray tracing program, developed by J.
Wambsganss \citep{Wyithe02b}, or  some analytical approach
for magnification near caustic curves like folds
\citep{Schneider92a,Fluke99} or near singular caustic points like
cusps \citep{Schneider92,Mao92,Zakharov95,Zakharov97,Zakharov99}
as was realized by \cite{Yonehara01}.

If  we believe in the  observational arguments of \cite{Hawk02} that the
variability of a significant fraction of distant quasars is caused by
microlensing,  the analysis of the properties of X-ray line shapes
due to microlensing \citep{Popovic03} is a powerful tool to
confirm or rule out Hawkins' (2002) conclusions.

As it was mentioned, the probability that the shape of the  Fe $K\alpha$
line is distorted (or amplified) is highest in gravitationally lensed
systems.
 Actually, this phenomena was
discovered by \cite{Osh01,Dai03,Chart02a,Cha02b,Cha04} who found
evidence for such an effect for   QSO H1413+117 (the Cloverleaf,
$z=2.56$), QSO~2237+0305 (the Einstein Cross, $z=1.695$), MG
J0414+0534 ($z=2.64$) and possibly for BAL QSO~08279+5255 ($
z=3.91$). Let us consider quasars located at the same redshifts as
the gravitational lensed objects. The probabilities that these
quasars are gravitationally microlensed by objects in a foreground
galaxy ($\tau_{GL}$) and by cosmologically distributed objects
($\tau^p_{L}$) are given in Table \ref{tabl2} (if we have no {\it
a priori} information about gravitational macrolensing for the
quasars). One can see from Table 2 that the optical depth for
microlensing by cosmologically distributed microlenses are one
order higher than for microlensing by  objects in a foreground
galaxy. So the observed microlensing in the X-ray Fe K$\alpha$
line from these objects should be caused by cosmologically
distributed objects rather than by the objects from a lensed
galaxy. For example,  in the case of the redshift corresponding to
the famous Einstein Cross QSO~2237+0305 where the optical depth is
smaller than for  other two redshifts.
 One could say that it is
natural that the discovery of X-ray microlensing was made for this
quasar, since the Einstein Cross QSO~2237+0305 is the
most "popular" object to search for microlensing, because the
first cosmological microlensing  phenomenon was found by
\cite{Irwin89} in this object and  several groups have been
monitoring the quasar QSO~2237+0305 to find evidence for
microlensing.  Microlensing has been suggested for the quasar
MG J0414+0534 \citep{Angonin99} and for the quasar QSO H1413+117
\citep{Remy96,Ostensen97,Turnshek97,Chae01}. Therefore,
in future may be a chance to find X-ray microlensing for other
gravitationally lensed systems that have  signatures of
microlensing in the optical and radio bands. Moreover, considering the
sizes of the sources of
X-ray radiation, the variability in the X-ray range during microlensing
event
should be more prominent than in the optical and UV.
{ Consequently, gravitational microlensing in the X-ray band
 is a powerful tool for
 dark matter investigations, as the upper limit of optical
depth ($\tau\sim 0.1$) corresponds to the case where dark matter forms
cosmologically distributed deflectors. On the other hand, one can see from
 Table 2 that, if we have no {\it a priori} information
about  gravitational lensing of distant quasars,  the expected
variabilities in the X-ray band due to microlensing tend to
be the same for the lensed and unlensed QSOs at the same
redshift. This means that cosmologically distributed deflectors play the
main
role in microlensing of high redshifted QSOs.  The comparison of X-ray
variation in lensed and unlensed QSOs at the same redshift can provide a
powerful test for the cosmologically distribution of the dark compact
objects.
 The observed rate of
 microlensing can be used for estimates of the cosmological density of
 microlenses (see, for example, Sect. 2.3), but  durations of
 microlensing events could be used to estimate microlens masses
 \citep{Wamb01,Wamb01b}.}

\begin{table}
\begin{center}
\caption[]{The calculated optical depths ($\tau_{GL}$ and
$\tau^p_L$ for 3 gravitational lensed objects. The used parameters
are: $\Omega_0=0.3,\ \Omega_L=0.05,\ F=0.15$. $\tau_{GL}$ is the
optical depth for macrolensing for quasars located at the same
redshifts as the gravitational lensed objects.} \label{tabl2}
\begin{tabular}{|c|c|c|c|}
\hline
Object               & z     & $\tau_{GL}$&$\tau^p_L$ \\
\hline \hline
MG J0414+0534        & 2.64  & 0.013652&0.1151256    \\
QSO 2237+0305        & 1.695 & 0.006635&0.0626277    \\
BAL QSO H1413+117 AT & 2.56  & 0.013049&0.1112457    \\
\hline
\end{tabular}
\end{center}
\end{table}

\section{Conclusions}

For a discussion of the contribution of microlensing to the X-ray
variability of high-redshift QSOs we calculated optical depth
considering the density of deflectors in the halo and bulge of
the host galaxy as well as for a cosmological distribution of
microdeflectors.

From our calculations we can conclude:

i) The optical depth in the bulge and halo of host galaxy is $\sim
10^{-4}$. This is in good agreement with previous estimates by
\cite{Popovic03}.
Microlensing by  deflectors from the host galaxy halo and
bulge makes a minor contribution to the  X-ray
variability of QSOs.

ii) The optical depth for  cosmologically distributed deflectors could
 be $\sim 10^{-2}-0.1$ at $z\sim 2$ and might contribute  significantly
 to the
X-ray variability of high-redshift QSOs. The value $\tau\sim 0.1$
corresponds to the
case where compact dark matter forms cosmologically distributed
microlenses.

iii) The optical depth for  cosmologically distributed deflectors
($\tau_L^p$)  is higher for $z>2$ and increases slowly beyond $z=2$.
 This indicates that the contribution of microlensing on the
X-ray variability of QSOs with redshift $z>2$ may be significant,
and also  that this contribution could  be nearly constant for
high-redshift QSOs. This is in good agreement with the fact that
AGNs of the same X-ray luminosity are more variable at $z>2$
\citep{Man02}.

{
 iv) Observations of X-ray variations of unlensed QSOs can be used for estimations of matter fraction of microlenses.
 The rate of
 microlensing can be used for estimates of the cosmological density of
microlenses, and consequently (see Sec. 2.3) the fraction of dark
matter microlenses, but the durations of microlensing events could be used
for gravitational microlens mass estimations.}

\begin{acknowledgements}

This work was supported by the National Natural Science Foundation
of China (No.:10233050) (AFZ), the Ministry of Science,
Technologies and Development of Serbia through the project
"Astrophysical Spectroscopy of Extragalactic Objects" (L\v CP \&
PJ) and the Alexander von Humboldt Foundation through the program for
foreign scholars (L\v CP).

AFZ is grateful to E.F.~Za\-kha\-rova for the kindness and support
necessary to complete this work. AFZ would like to thank the
National Astronomical Observatories of the Chinese Academy of
Sciences  for their hospitality and prof.~J.Wang and Dr.~
Z.Ma for very
useful discussions.

The authors  are grateful to an anonymous referee for very useful
remarks.
\end{acknowledgements}


\begin{thebibliography}{}

\bibitem[\protect\citeauthoryear{Abajas et al.}{2002}]{Aba02}
Abajas, C., Mediavilla, E.G., Mu\~noz, J.A., Popovi\'c, L. \v C.,
\& Oscoz A. 2002,  ApJ 576, 640.

\bibitem[\protect\citeauthoryear{Angonin-Willaime et al.}{1999}]{Angonin99}
 Angonin-Willaime, M.-C., Vanderriest, C., Courbin, F., et al.
1999, A\&A, 347, 434.


\bibitem[\protect\citeauthoryear{Aretxaga \& Terlevich}{1994}]{Aret94}
Aretxaga, I., \& Terlevich, R. 1994,  MNRAS, 269, 462

\bibitem[\protect\citeauthoryear{Balbi}{2001}]{Bal01}
Balbi, A. 2001, in  Cosmology and particle physics, ed.  R. Durrer
et al.,  AIP Conference Proceedings,  555, 107

\bibitem[\protect\citeauthoryear{Barr, Mushotzky}{1986}]{Bar86}
Barr, P., \& Mushotzky, R. F. 1986,  Nature, 320, 421

\bibitem[\protect\citeauthoryear{Bennett et al.}{2003}]{Ben03}
 Bennett, C.L.,  Halpern, M., Hinshaw, G., et al., 2003, accepted in ApJ,
astro-ph/0302207

\bibitem[\protect\citeauthoryear{Bian \& Zhao}{2003}]{Bian03}
Bian, W.-H., \& Zhao, Y.-H. 2003, Chin. J. Astron. Astrophys., 3,
119

\bibitem[\protect\citeauthoryear{Bond et al.}{2001}]{Bon01}
Bond, J.R., Pogosyan, D.,  Prunet, S., et al.
 2001, in  Cosmology and particle physics, ed.
R. Durrer et al.,  AIP Conference Proceedings,  555, 263

\bibitem[\protect\citeauthoryear{Bridle  et al.}{2003}]{Bridle03}
 Bridle, S.L., Lahav, O., Ostriker, J.P., \& Steinhardt, P.J. 2003,
 Science, 299, 1532

 \bibitem[\protect\citeauthoryear{Browne et al.}{2003}]{Browne03}
 Browne, I.W.A., Wilkinson, P.N., Jackson, N.J.F.,
et al., 2003,
MNRAS, 341, 13

\bibitem[\protect\citeauthoryear{Bullock et al.}{2001}]{Bullock01}
Bullock, J.S., Kolatt, T.S., Sigad, Y., et al., 2001, MNRAS, 321,
559

\bibitem[\protect\citeauthoryear{Burles, Nollett \& Turner}{2001}]{Burles01}
Burles, S., Nollett, K., \& Turner, M.S.,  2001,  ApJ,  552, L1

\bibitem[\protect\citeauthoryear{Chae  et al.}{2001}]{Chae01}
 Chae, K.-H., Turnshek, D.A., Schulte-Ladbeck, R.E., et al.
2001, ApJ,  568, 509

\bibitem[\protect\citeauthoryear{Chartas et al.}{2002a}]{Chart02a}
Chartas, G., Agol, E., Eracleous, M., et al.
2002a,  ApJ,  568, 509

\bibitem[\protect\citeauthoryear{Chartas et al.}{2002b}]{Cha02b}
Chartas, G., Brandt, W.N., Gallagher, S.M., \& Garmire, G.P. 2002b,
ApJ, 579, 169

\bibitem[\protect\citeauthoryear{Chartas et al.}{2004}]{Cha04}
Chartas, G., Eracleous, M., Agol, E., Gallagher, S.C. 2004, ApJ accepted
(astro-ph/0401240).

\bibitem[\protect\citeauthoryear{Claeskens \& Surdej}{2002}]{Claeskens02}
Claeskens, J.F., \& Surdej, J. 2002, Astron \& Astroph. Rev., 10,
263

\bibitem[\protect\citeauthoryear{Czerny et al.}{2001}]{Czerny01}
Czerny, B., Nikolajuk, M., Piasecki, M., \& Kuraszkiewicz, J.,
2001, MNRAS, 325, 865

\bibitem[\protect\citeauthoryear{Dai et al. }{2003}]{Dai03}
Dai, X., Chartas, G., Agol, E., Bautz, M. W., \&
  Garmire, G.P. 2003, ApJ, 589, 100

\bibitem[\protect\citeauthoryear{Deguchi \& Watson}{1987}]{Deguchi87}
Deguchi, S., \& Watson, W.D. 1987, Phys. Rev. Lett., 59(24), 2814

\bibitem[\protect\citeauthoryear{Ferrarese}{2002}]{Ferrarese02}
Ferrarese, L. 2002, ApJ, 578, 90

\bibitem[\protect\citeauthoryear{Fluke \& Webster}{1999}]{Fluke99}
Fluke, C.J., \& Webster, R.L. 1999, MNRAS,  302, 68

\bibitem[\protect\citeauthoryear{Freedman et al.}{2001}]{Freedman01}
 Freedman, W.L., Madore, B.F., Gibson, B.K., et al.,
2001, ApJ, 553, 47

\bibitem[\protect\citeauthoryear{Fukugita and Turner}{1991}]{Fuk91}
Fukugita, M., \& Turner, E.L. 1991,  MNRAS,  253, 99

\bibitem[\protect\citeauthoryear{Irwin et al.}{1989}]{Irwin89}
Irwin, M.J., Webster, R.L., Hewett, P.C., et al.,
1989, AJ, 98, 1989

\bibitem[\protect\citeauthoryear{Hawkins}{1993}]{Hawk93}
Hawkins, M.R.S. 1993,  Nature,  366, 242

\bibitem[\protect\citeauthoryear{Hawkins}{1996}]{Hawk96}
Hawkins, M.R.S. 1996, MNRAS,  278, 787

\bibitem[\protect\citeauthoryear{Hawkins}{2002}]{Hawk02}
Hawkins, M.R.S. 2002,  MNRAS,  329, 76

\bibitem[\protect\citeauthoryear{Gott}{1981}]{Gott81}
Gott, J.R. 1981, ApJ, 243, 140

\bibitem[\protect\citeauthoryear{Green et al.}{1993}]{Gre93}
Green, A.R., McHardy, I.M., \&  Lehto, H.J. 1993,  MNRAS,  265,
664

\bibitem[\protect\citeauthoryear{Kawaguchi et al.}{1998}]{Kawa98}
Kawaguchi, T., Mineshige, S., Umemura, M., \& Turner, E.L. 1998,
ApJ, 504, 671

\bibitem[\protect\citeauthoryear{Klypin et al.}{2002}]{Klypin02}
Klypin, A.A., Zhao, H., \& Somerville, R.S. 2002, ApJ, 573, 597

\bibitem[\protect\citeauthoryear{Lahav}{2002}]{Lahav02}
Lahav, O. 2002, astro-ph/0208297

\bibitem[\protect\citeauthoryear{Lawrence \& Papadakis}{1993}]{Law93}
Lawrence, A., \& Papadakis, I.  1993,  ApJ,  414, 85

\bibitem[\protect\citeauthoryear{Manners et al.}{2002}]{Man02}
Manners, J., Almaini, O., \& Lawrence, A. 2002,  MNRAS,  330, 390

\bibitem[\protect\citeauthoryear{Marshall et al.}{1981}]{Mar81}
Marshall, N., Warwick, R.S., Pounds, K.A. 1981,  MNRAS,  194, 987

\bibitem[\protect\citeauthoryear{Mao}{1992}]{Mao92}
Mao, S.  1992,  ApJ,  389, 63

\bibitem[\protect\citeauthoryear{McLure \& Dunlop}{2002}]{Mclure02}
McLure, R.J., \& Dunlop, J.S. 2002,  MNRAS, 331, 795

\bibitem[\protect\citeauthoryear{Mushotzky et al.}{1993}]{Mush93}
Mushotzky, R.F., Done, C. \& Pounds, K. A. 1993, ARA\&A, 31, 717

\bibitem[\protect\citeauthoryear{Navarro et al.}{1995}]{NFW95}
Navarro, J.F., Frenk, C.S., \& White, S.D.M. 1995, MNRAS, 275, 56

\bibitem[\protect\citeauthoryear{Navarro et al.}{1996}]{NFW96}
Navarro, J.F., Frenk, C.S., \& White, S.D.M. 1996, MNRAS, 462, 563

\bibitem[\protect\citeauthoryear{Navarro et al.}{1997}]{NFW97}
Navarro, J.F., Frenk, C.S., \& White, S.D.M. 1997, MNRAS, 490, 493

\bibitem[\protect\citeauthoryear{Neindorf}{2003}]{Neindorf03}
Neindorf, B. 2003, A \& A, 404, 83

\bibitem[\protect\citeauthoryear{O'Meara et al.}{2001}]{Omeara01}
 O'Meara, J. M., Tytler, D., Kirkman, D., et al.
2001, ApJ, 552, 718

\bibitem[\protect\citeauthoryear{Oshima et al.}{2002}]{Osh01}
Oshima, T., Mitsuda, K., Fujimoto R., Iyomoto N., Futamoto K., et
al., 2001,   ApJ,  563, L103

\bibitem[\protect\citeauthoryear{Ostensen et al.}{1997}]{Ostensen97}
 Ostensen, R., Remy, M., Lindblad, P.O., et al.,
1997, A\&AS, 126, 393

\bibitem[\protect\citeauthoryear{Parodi et al.}{2000}]{Parodi00}
Parodi, B.R., Saha, A., Sandage, A., \& Tammann, G.A.
2000,  ApJ, 540, 634

\bibitem[\protect\citeauthoryear{Peebles}{2002}]{Peebles02}
Peebles, P.J.E. 2002, astro-ph/0208037

\bibitem[\protect\citeauthoryear{Perlmutter et al.}{1999}]{Per99}
 Perlmutter, S., Aldering, G., Goldhaber, G., et al., 1999,
ApJ, 517, 565

\bibitem[\protect\citeauthoryear{Popovi{\'c} et al.}{2001a}]{Popov01a}
Popovi{\'c}, L., \v C.,  Mediavilla, E.G., Mu\~noz J.,
Dimitrijevi\'c, M.S., \& Jovanovi\'c, P. 2001a,  Serb. Aston. J.,
164, 73. (Also, presented on GLITP Workshop on Gravitational Lens
Monitoring, 4-6 June 2001, La Laguna, Tenerife, Spain)

\bibitem[\protect\citeauthoryear{Popovi{\'c} et al.}{2001b}]{Pop01b}
Popovi{\'c}, L.\v C.,  Mediavilla, E.G., \& Mu\~noz J., 2001b,
A{\&}A 378, 295.

\bibitem[\protect\citeauthoryear{Popovi{\'c} et al.}{2003}]{Popovic03}
 Popovi\'c, L.\v C., Mediavilla, E.G., Jovanovi\'c, P., \& Mu\~noz, J.A.,
 2003,  A \&  A, 398, 975

\bibitem[\protect\citeauthoryear{Rees}{1984}]{Rees84}
Rees, M., 1984,  ARA \& A, 22, 471

\bibitem[\protect\citeauthoryear{Remy et al.}{1996}]{Remy96}
 Remy, M., Gosset, E., Hutsemekers, D., et al. 1996,
in Astrophysical
applications of gravitational lensing:
 Proceedings of the 173rd IAU Symposium,
ed.  C. S. Kochanek and J.N. Hewitt,
(Kluwer Academic Publishers; Dordrecht) 261

\bibitem[\protect\citeauthoryear{Schade et al.}{2000}]{Schade00} Schade,
D.J., Boyle, B.J., \& Letawsky,
M. 2000,  MNRAS,  194, 987

\bibitem[\protect\citeauthoryear{Schneider}{1992}]{Schneider92a}
Schneider, P. 1992, Gravitational Lenses, (Springer, Berlin)

\bibitem[\protect\citeauthoryear{Schneider \& Weiss}{1992}]{Schneider92}
 Schneider, P., \& Weiss, A. 1992, A \& A, 260, 1

\bibitem[\protect\citeauthoryear{Seitz \& Schneider}{1994}]{Seitz94a}
Seitz, C., \& Schneider, P. 1994, A \& A, 288, 1

\bibitem[\protect\citeauthoryear{Seitz et al.}{1994}]{Seitz94}
Seitz, C., Wambsganss, J., \& Schneider, P. 1994, A \& A, 335, 379

\bibitem[\protect\citeauthoryear{Spergel et al.}{2003}]{Sper03}
 Spergel, D.N.,  Verde, L.,  Peiris, H.V.,    et al.
  2003, ApJS, 148, 175

\bibitem[\protect\citeauthoryear{Shields et al.}{2003}]{Shields02}
Shields, G.A., Gebhard, K., Salviander, S., et al. 2003, ApJ, 583,
124

\bibitem[\protect\citeauthoryear{Stompor et al.}{2001}]{Stompor01}
 Stompor, R., Abroe, M., Ade, P., et al.
2001, ApJ, 561, L7

\bibitem[\protect\citeauthoryear{Takahashi et al.}{2001}]{Tak01}
Takahashi, R., Yonehara, A., \& Mineshige, S. 2001,  PASJ, 53, 387

\bibitem[\protect\citeauthoryear{Tammann \& Reindl}{2002a}]{Tammann02}
 Tammann, G.A., \& Reindl, B.,
2002a,  Ap \& SS, 280, 165

\bibitem[\protect\citeauthoryear{Tammann \& Reindl}{2002b}]{Tammann02a}
Tammann, G.A., \& Reindl, B., 2002b,
 astro-ph/0208176

\bibitem[\protect\citeauthoryear{Turner}{1990}]{Turner90}
Turner, E.L. 1990,  ApJ,  365, L43

\bibitem[\protect\citeauthoryear{Turner et al.}{1984}]{TOG84}
Turner, E.L., Ostriker, J.P., \& Gott, J.R. 1984,  ApJ,  284, 1

\bibitem[\protect\citeauthoryear{Turner}{2002}]{Turner02}
Turner, M.S.  2002,  ApJ,  576, L101

\bibitem[\protect\citeauthoryear{Turner et al.}{1999}]{Tur99}
Turner, T.J., George, I.M., Nandra, K., \& Turcan, D. 1999, ApJ,
524, 667

\bibitem[\protect\citeauthoryear{Turnshek et al.}{1997}]{Turnshek97}
 Turnshek, D.A., Lupie, O.L., Rao, S.M., et al.,
1997, ApJ, 485, 100

\bibitem[\protect\citeauthoryear{Ulrich et al.}{1993}]{Ulrich97}
Ulrich, M.-H., Maraschi, L., \& Megan, C.M. 1997, ARA\&A, 35, 445

\bibitem[\protect\citeauthoryear{Walsh et al.}{1979}]{Walsh79}
Walsh, D., Carswell, R.F., \&  Weymann, R.J. 1979, Nature, 279,
381

\bibitem[\protect\citeauthoryear{Wambsganss}{2001a}]{Wamb01}
   Wambsganss, J. 2001a,
 in Microlensing 2000: A new Era of Microlensing Astrophysics,
ed. J.W.Menzies and P.D.Sackett
 ASP Conf. Series, 239, 351

\bibitem[\protect\citeauthoryear{Wambsganss}{2001b}]{Wamb01b}
   Wambsganss, J. 2001b,  PASA, 18, 207

\bibitem[\protect\citeauthoryear{Wang, Stebbins \& Turner}{1996}]{Wan96}
Wang, Y., Stebbins, A., \&  Turner, E.L. 1996, Phys. Rev. Lett.,
77, 2875

\bibitem[\protect\citeauthoryear{Weaver et al.}{2001}]{Wea01}
Weaver, K.A., Gelbord, J., \& Yaqoob, T. 2001,  ApJ,  550, 261

\bibitem[\protect\citeauthoryear{Wyithe \& Turner}{2002a}]{Wyithe02}
Wyithe, J.S.B., \& Turner, E.L. 2002a, ApJ, 567, 18

\bibitem[\protect\citeauthoryear{Wyithe \& Turner}{2002b}]{Wyithe02b}
Wyithe, J.S.B. \& Turner, E.L. 2002b, ApJ, 575, 650

\bibitem[\protect\citeauthoryear{Yonehara}{2001}]{Yonehara01}
   Yonehara, A. 2001,  PASA, 18, 211

\bibitem[\protect\citeauthoryear{Zakharov}{1995}]{Zakharov95}
 Zakharov, A.F. 1995,  A \&  A, 293, 1

\bibitem[\protect\citeauthoryear{Zakharov}{1997a}]{Zakh97}
Zakharov, A.F. 1997a,  Gravitational lenses and microlensing,
(Janus-K, Moscow)

\bibitem[\protect\citeauthoryear{Zakharov}{1997b}]{Zakharov97}
 Zakharov, A.F. 1997b,  Ap\&SS, 252, 369

\bibitem[\protect\citeauthoryear{Zakharov}{1999}]{Zakharov99}
Zakharov, A.F. 1999,
in  Recent Developments in Theoretical and Experimental General
Relativity, Gravitation, and Relativistic Field Theories,  ed.
Ts.~Piran and R.~Ruffini (World Scientific Publishers, Singapore),
1500

\bibitem[\protect\citeauthoryear{Zakharov et al.}{2003}]{ZKLR02}
   Zakharov, A.F.,  Kardashev, N.S., Lukash, V.N., \& Repin, S.V. 2003,
MNRAS, 342, 1325

\bibitem[\protect\citeauthoryear{Zakharov \& Repin}{2002a}]{Zak_rep02}
   Zakharov, A.F., \& Repin, S.V. 2002a, Astronomy Reports, 46, 360

\bibitem[\protect\citeauthoryear{Zakharov \& Repin}{2002b}]{Zak_rep02a}
   Zakharov, A.F., \& Repin, S.V. 2002b,
 in Proc. of the Eleven Workshop
   on General Relativity  and Gravitation in Japan, ed.
   J.~Koga, T.~Nakamura, K.~Maeda, K.~Tomita, (Waseda University,
   Tokyo) 68

\bibitem[\protect\citeauthoryear{Zakharov \& Repin}{2002c}]{Zak_rep02_xeus}
   Zakharov, A.F., \& Repin, S.V. 2002c, in
   Proc. of the Workshop "XEUS - studying the evolution of the hot
   Universe", ed. G. Hasinger, Th. Boller, A.N. Parmar,  MPE Report~281, 339

\bibitem[\protect\citeauthoryear{Zakharov \& Repin}{2003a}]{Zak_rep03_ASR}
   Zakharov, A.F., \& Repin, S.V. 2003a, accepted in Advances in Space
Res.


\bibitem[\protect\citeauthoryear{Zakharov \& Repin}{2003b}]{Zak_rep03_AA}
   Zakharov, A.F., \& Repin, S.V. 2003b, 406,7

\bibitem[\protect\citeauthoryear{Zakharov \& Repin}{2003c}]{Zak_rep03_Su}
   Zakharov, A.F., \& Repin, S.V. 2003c, in
   Proc. of the 214th IAU Symposium "High Energy Processes and Phenomena in
   Astrophysics", ed. X.D. Li, V. Trimble, Z.R. Wang,  vol. 214, 97

\end{thebibliography}
\end{document}